\documentclass[a4paper,12pt]{article}
\usepackage{graphicx,amssymb,color}
\usepackage{epsfig}
\usepackage{color}

\begin{document}

\centerline{\LARGE \bf  Immuno-epidemiological model of}

\medskip

\centerline{\LARGE \bf    two-stage epidemic growth}

\vspace*{1cm}

\centerline{\bf Malay Banerjee$^1$, Alexey Tokarev$^2$,  Vitaly Volpert$^{3, 4, 2}$ }

\vspace*{0.5cm}

\centerline{$^1$ Department of Mathematics \& Statistics, IIT Kanpur, Kanpur - 208016, India}

\centerline{$^2$
Peoples’ Friendship University of Russia (RUDN University)}
\centerline{ 6 Miklukho-Maklaya St, Moscow, 117198, Russia}

\centerline{$^3$
Institut Camille Jordan, UMR 5208 CNRS, University Lyon 1, 69622 Villeurbanne, France}

\centerline{$^4$
INRIA
Team Dracula, INRIA Lyon La Doua, 69603 Villeurbanne, France}

\vspace*{1cm}

\noindent
{\bf Abstract.}
Epidemiological data on seasonal influenza show that the growth rate of the number of infected individuals can increase passing from one exponential growth rate to another one with a larger exponent. Such behavior is not described by conventional epidemiological models. In this work an immuno-epidemiological model is proposed in order to describe this two-stage growth. It takes into account that the growth in the number of infected individuals increases the initial viral load and provides a passage from the first stage of epidemic where only people with weak immune response are infected to the second stage where people with strong immune response are also infected.
This scenario may be viewed as an increase of the effective number of susceptible increasing the effective growth rate of infected.

\vspace*{0.5cm}


\vspace*{0.25cm}


\vspace*{0.25cm}


\section{Introduction}

On March 22, 2020 according to the official data presented at the Worldometer \cite{W} for COVID-19, the number of infected individuals exceeded 10 thousand in seven countries. The respective growth rate of the number of infected is shown in Table 1.

The first column represents the number of days needed to pass from 100 to 1000 infected, the second column from 1000 to 10000. If the number of days  is the same in the two columns, then the growth is exponential, as it is predicted by conventional epidemiological models (SIR) for the beginning of epidemic when the number of uninfected (susceptible) individuals can be considered as approximately constant. In Italy, Iran, and France, the number of days in the second column is larger than in the first column. In Italy main restrictions were adopted beyond ten thousand initial infected. In Germany this difference is small at the moment, so it is early to make conclusions. In USA and Spain the number of days is the same and the growth rate is slightly larger in Spain.

\begin{center}
\begin{tabular}{|c|c|c|}
  \hline
   & 100-1000 & 1000-10000 \\
\hline
  China & - & 7 \\
  Italy & 6 & 10 \\
  USA   & 8 & 8 \\
  Spain & 7 & 7 \\
  Germany & 8 & 9 \\
  Iran    & 5 & 10 \\
  France  & 7 & 11 \\
  \hline
\end{tabular}
\end{center}
\noindent
{\bf Table 1.} Tenfold time of the number of infected individuals in the country where their total number exceeds 10 thousand on March 22. The first column shows the number of days required to pass from 100 to 1000 infected, the second column from 1000 to 10 000 infected.

\medskip
\medskip

\noindent
These data should be taken with caution because they can be strongly influenced by the number of effectuated tests. According to some media reports, there is a shortage of testing facility and/or shortage of testing kits in many countries, and the accurate number of infected individuals can be much larger \cite{lach}.

Dynamics of the number of infected individuals can be influenced by various factors, in particular, by their spatial distribution and nature of social mixing. Appearance  of new focuses of infection can essentially change the growth curves. For example, at the whole world level, development of the epidemic in Europe started when it was mostly at decaying mode in Asia. Instead of the standard growth-decay curves, this leads to growth-decay-growth curves. Some elements of this more complex dynamics where the number of newly reported daily cases decreased during several days was also observed in some countries.

If we suppose that the distribution of infected individuals is uniform over space, then the SIR model predicts the exponential growth at the beginning of epidemic. Some time later, the growth rate decreases due to the decrease in the number of susceptible individuals, and the number of infected individuals decays at the end of epidemic. There are recent works with various modifications of the epidemiological models trying to take into account some specific features of the coronavirus pandemic \cite{koch, liu, VBP}.

Analysis of the data on seasonal influenza shows that dynamics of epidemic spread can include the period of accelerated growth (Figure \ref{flu}). In the beginning of epidemic, the growth rate is exponential. At the next stage it continues to grow exponentially but with an increased growth rate. It can be considered as two different exponential functions where the second one has a larger exponent than the previous one. A possible explanation of this effect is related to the presence of sub-populations with different response on the infection. Simplifying the situation, we can consider two sub-populations, one of them with a weak immune response and another one with a strong immune response. The strength of immune response is determined by the rate of production of antigen specific immune cells. We will define it below. At the beginning of epidemic, infection propagates among the first sub-population. When the epidemic reaches certain level, the second sub-population becomes also involved increasing the number of susceptible individuals.

Transition between the two stages of epidemic growth can be related to the number of infected individuals. If it is small enough, then a healthy individual rarely meets infected individuals, and the initial viral load remain small. People with a strong immune system can eliminate infection without developing disease symptoms, while people with a weak immune system can fall sick. At the later stages of epidemic, when the number of infected individuals is sufficiently large, an uninfected individual can cross several infected individuals during a short period of time. Therefore, the initial viral load becomes larger, and people with a strong immune response can also become ill.

In order to describe this two stage epidemic growth, we develop in this work an immuno-epidemiological model combining the individual level and the population level. This model is generic but it can take into account specific features of a particular infection or the details of epidemiological situation (like quarantine, social distancing, self-isolation, etc.). In the next section, we will discuss some epidemiological data on influenza epidemic. Section 3 is devoted to a model of immune response where we show how people with weak and strong immune system can react on a given initial viral load. Immuno-epidemiological model is studied in Section 4.



\section{Two stage epidemic}

\begin{figure}[ht!]
\centerline{\includegraphics[scale=0.2]{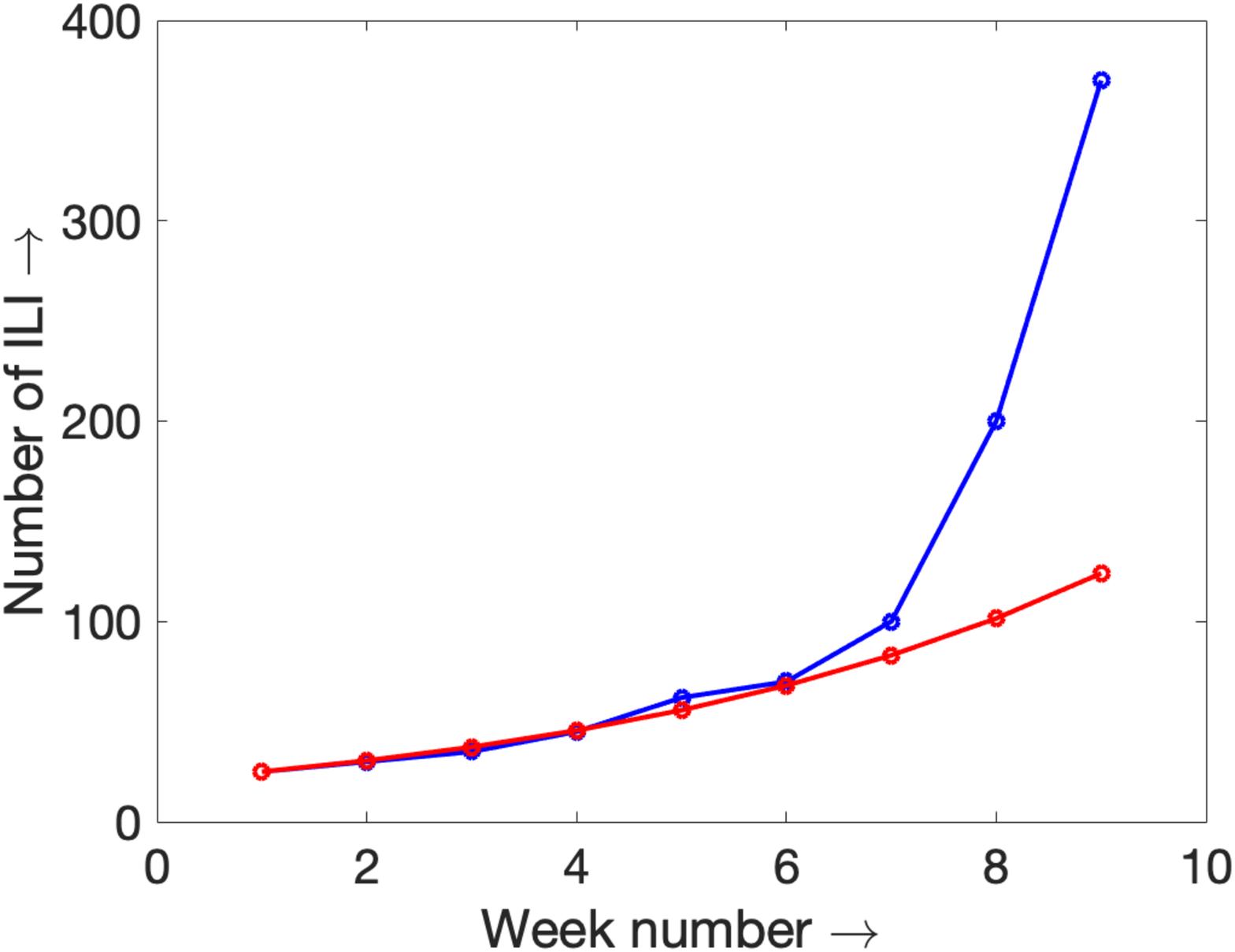}
\includegraphics[scale=0.2]{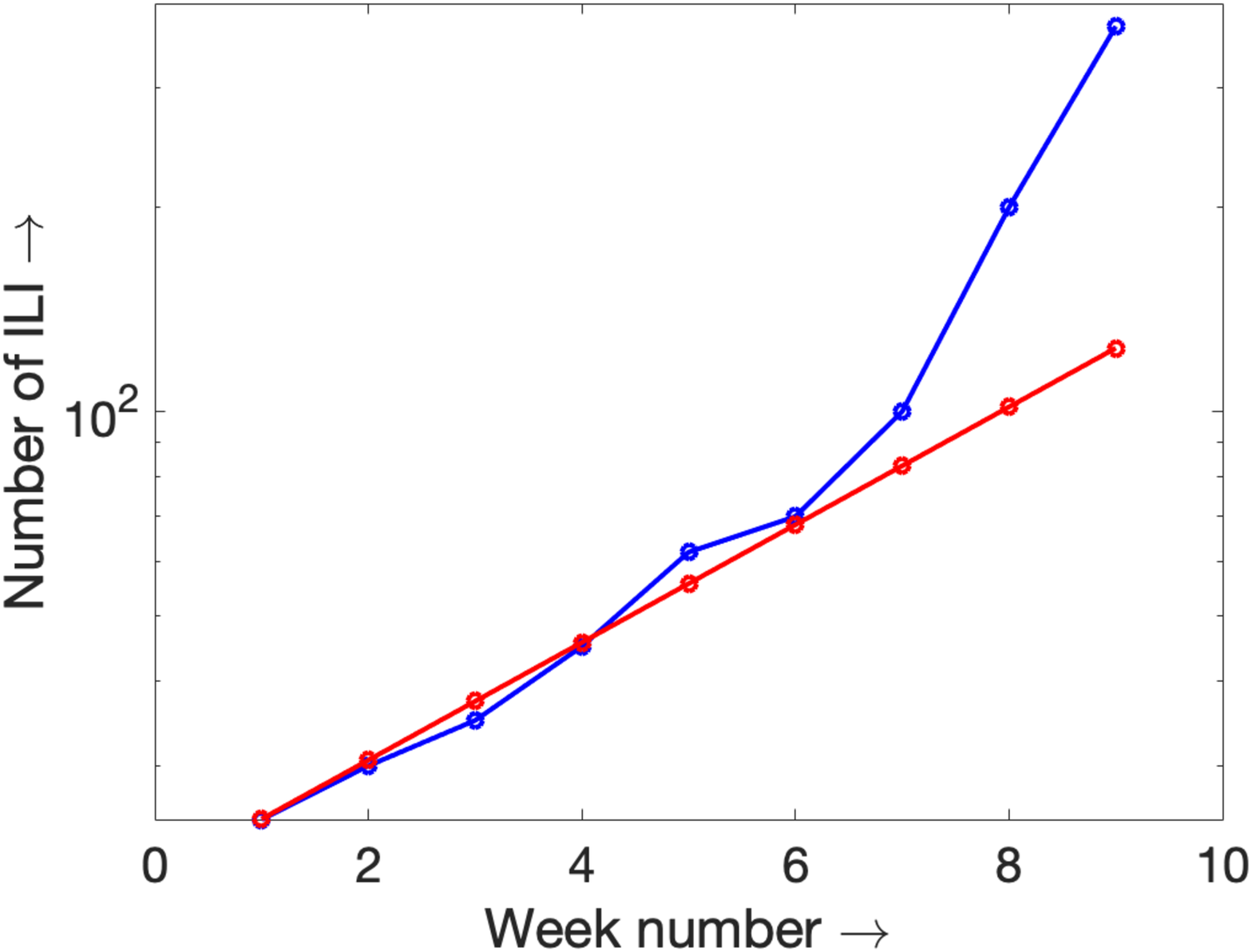}}
\caption{Data on seasonal influenza in Europe in 2019-2020 \cite{who}. The left graph shows the number of influenza-positive specimens from non-sentinel sources (blue line) and its approximation by an exponential (red line). The right graph shows the same curves in the logarithmic scale.}
\label{flu}
\end{figure}

There are large amount of data available for the seasonal influenza epidemic. A typical graph of the number of cases is shown in Figure \ref{flu} (left, blue curve) \cite{who}. Growth rate of the number of positive cases between weeks 1 and 6 is exponential. The red curve shows the approximation of the blue curve with the function $20.5 \exp(0.2 x)$, where $x$ is the number of weeks. The same curves are shown in Figure \ref{flu} (right) in the logarithmic scale. The red curve here is a straight line. The weeks 7-10 show different growth rate with some acceleration in growth. It also corresponds to an exponential growth but with a different exponent. Some other data are presented in Appendix 1.

Conventional epidemiological models like SIR (Susceptible, Infected, Recovered, see, e.g., \cite{mur}) and their numerous variants describe the density of infected individuals $I$ by the typical equation

\begin{equation}\label{a1}
  \frac{dI}{dt} = \lambda I S - \nu I ,
\end{equation}
where $S$ and $I$ are the densities of susceptible and infected individuals, and $\lambda$ is a positive constant. The first term in the right-hand side of this equation characterizes the appearance of new infected individuals due to their contact with susceptible. The second term corresponds to the decrease of $I$ due to the recovery or death of infected individuals.

In the beginning of the infection spreading, the number of infected and recovered individuals are much less than the number of susceptible, so that we can approximate $S$ by a constant $S \approx S_0$. Using this approximation, we obtain a linear differential equation with constant coefficients. Its solution is given by the function $I(t) = I_0 \exp(\alpha t)$, where $I_0$ is the number of infected individuals at the initial moment of time, and $\alpha = \lambda S_0-\nu$. If $\alpha > 0$, then $I(t)$ exponentially growth. The same condition can be written as $\mathcal{R}_0>1$, where $\mathcal{R}_0=\lambda S_0/\nu$ is the basic reproduction number.

Thus, according to equation (\ref{a1}), growth of the number of infected individuals is exponential in the beginning of epidemic, and it slows down later when $S$ decreases. Exponential growth in the beginning corresponds to the data in Figure \ref{flu}. However, it does not describe accelerated growth during the second stage of epidemic. A possible explanation of this acceleration is that in the beginning the epidemic, disease spreads mainly among the people with weak immune response. At the second stage, people with strong immune response become also exposed. In order to investigate this hypothesis, we consider a model of immune response in the next section.


\bigskip

\setcounter{equation}{0}

\section{Model of immune response}

We consider a qualitative model of immune response suggested in \cite{BMBTV}:
\begin{equation}\label{b1}
  \frac{dv}{dt} = k v (1-v) - \sigma c v ,
\end{equation}
\begin{equation}\label{b2}
  \frac{dc}{dt} = p(v) c (1-c) - h(v) c ,
\end{equation}
where $v$ and $c$ are the concentrations of viruses and immune cells within the organism. The first term in the right-hand side of equation (\ref{b1}) describes the virus multiplication rate. This logistic term is proportional to the virus density $v$ and to the normalized density of uninfected cells $(1-v)$. The second term characterizes virus death due to the immune response. This term is proportional to the virus density and to the density of immune cells $c$. The first term in the right-hand side of equation (\ref{b2}) describes multiplication of immune cells and the second term their mortality. In general, both of them depend on virus concentration. Clonal expansion of immune cells is stimulated by antigen (virus) if its concentration is not very large. For large virus concentration, the proliferation rate of immune cells can decrease. Therefore, the function $p(v)$ is non-negative and increasing with saturation or growing for small $v$ and decreasing for large $v$. The death rate of immune cells can also depend on the concentration of viruses. In this case it is a positive increasing function. It can be also considered as constant if this effect is negligible.

\begin{figure}[ht!]
\centerline{\includegraphics[scale=0.8]{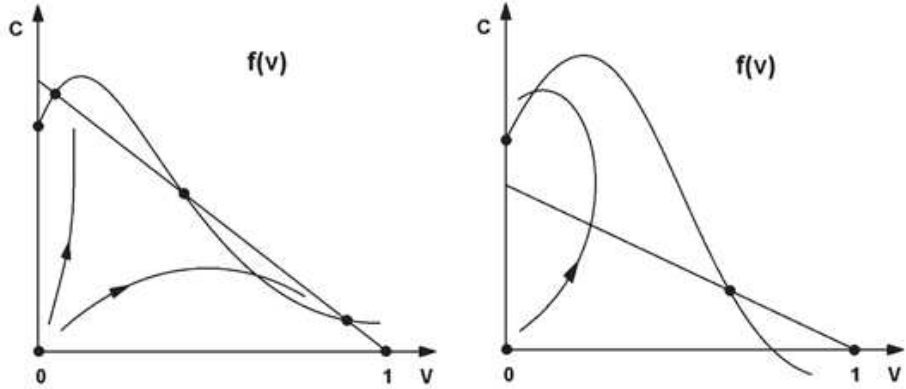}}
\caption{Two examples of phase portrait of system (\ref{b1}), (\ref{b2}) depending on the function $f(v) = 1 - h(v)/p(v)$. It can have six (left) or four (right) stationary points. The trajectory with the initial condition $(v_0,c_0)$ and sufficiently small viral load $v_0$ can go either to the stationary point with complete virus elimination (right) or to a weak persistent infection (left). If the initial viral load is sufficiently large, then the trajectory approaches the stationary point with large $v$ and small $c$. It corresponds to a strong chronic infection or to death. Reprinted from \cite{BMBTV}.}
\label{ode}
\end{figure}

The nullclines of system (\ref{b1}), (\ref{b2}) are given by the expressions: $v=0$ and $c=k(1-v)/\sigma$; $c=0$ and $c=f(v)$, where $f(v) = 1 - h(v)/p(v)$. Figure \ref{ode} qualitatively shows possible phase portraits of this system. There are six stationary points in the left phase portrait and four in the right one. If there is a small initial virus load $v_0$ and a small number of $c_0$ for antigen specific immune cells (new infection, no memory cells), then the trajectory of this system will start from a small vicinity of the unstable stationary point $(0,0)$ and move towards one of the other stationary points. Summarizing the behavior of solutions, we observe that if the initial viral load $v_0$ is sufficiently small, then the trajectory approaches the stationary point with complete curing (Figure \ref{ode}, right) or with a weak persistent infection (Figure \ref{ode}, left). If the initial viral load is large enough, then the solution converges to a stationary point with a large virus concentration and low concentration of immune cells. It can characterize a strong chronic infection or a lethal outcome. We present some results of numerical simulations in Appendix 2.

In order to get an explicit solution, we simplify system (\ref{b1}), (\ref{b2}) with the following assumptions:

\medskip
\noindent
(A1) - functions $p(v)$ and $h(v)$ are constant. This means that the clonal reproduction of immune cells fully responds to the antigen as soon as there is a small amount of virus in the organism, and mortality rate of the immune cells does not depend on virus concentration;

\medskip
\noindent
(A2) - there is no limitation on the production of viruses and immune cells by available resources, i.e., the logistic terms in equations (\ref{b1}), (\ref{b2}) are replaced by the linear terms.

\medskip
\noindent
Under these assumptions, instead of system (\ref{b1}), (\ref{b2}) we consider the following system of equations:
\begin{equation}\label{b3}
  \frac{dv}{dt} = k v  - \sigma c v ,
\end{equation}
\begin{equation}\label{b4}
  \frac{dc}{dt} = p c  - h c .
\end{equation}
We find from equation (\ref{b4}), $c=c_0 e^{\mu t}$, where $c_0$ is the number of antigen specific immune cells initially, and $\mu = p-h$. Assuming that $\mu>0$, we obtain exponential growth of immune cells. Substituting the solution for $c$ into equation (\ref{b3}), we find the solution for $v$ as
$$ v(t) = v_0 e^{kt - \frac{c_0 \sigma}{\mu} (e^{\mu t} - 1)} \; , $$
where $v_0$ is the initial viral load. The derivative of this function
$$ v'(t) = v_0  \left(k - c_0 \sigma e^{\mu t} \right)    e^{kt - \frac{c_0 \sigma}{\mu} (e^{\mu t} - 1)} $$
is positive at $t=0$ since the initial number $c_0$ of antigen specific cells is very small. Therefore, $v(t)$ grows at the beginning of infection development. This derivative equals $0$ at time $t_* = \frac{1}{\mu}\ln\left(\frac{k}{c_0\sigma}\right)$ when the maximal virus concentration is reached. After that, it decreases and converges to $0$. The maximal virus concentration is as follows:
$$ v^* = v_0 e^{(w \ln w - w + 1) /s} \; , $$
where $w=k/(c_0\sigma)$, $s=(p-h)/(c_0\sigma)$. Parameter $w$ corresponds to the dimensionless virus multiplication rate. It characterizes the virulence of infection. Parameter $s$ characterizes the strength of the immune response.

The maximal level of viral concentration is determined by the initial viral load $v_0$, the virulence of infection $w$, and the strength of immune response $s$. For a given virulence of infection $w>1$, the maximal viral load increases with for larger initial viral loads and decreases for stronger immune response. If we set $W = w \ln w - w + 1$, then $v^* = v_0 \exp(W/s)$.

Suppose that there are two levels of immune response $s_1$ and $s_2$,  with $s_1 < s_2$. The first one corresponds to weak immune response and $s_2$ to strong immune response. For a given initial viral load $v_0$, the corresponding maximal virus concentrations are $v_1^*$ and $v_2^*$, where $v_1^* > v_2^*$. Disease development depends on virus concentration in a threshold way. If its level exceeds some critical value, which is determined by a proportion of infected tissue, then disease symptoms are developed, and the person is considered as infected and fall sick. If this critical level of virus concentration $v_c$ is such that $v_1^* > v_c > v_2^*$, then people with weak immune response become ill while people with strong immune response remain healthy. The same property is preserved in the full model, where $s_1<s_2$ leads to $v_1^*>v_2^*$ for the same $v_0$ (see Appendix 2).

However, if the initial virus load increases with the progression of epidemic, then $v_2^*$ also increases and it can overpass the critical level. We consider this question in the next section.


\setcounter{equation}{0}

\section{Immuno-epidemiological model}

We can now bring together the epidemiological model and the immune response model making a link between the number of infected individuals $I$, the initial viral load $v_0$, and the number of susceptible individuals $S$. We suppose that the $v_0=v_0(I)$ and $S\approx S(v_0)= S(v_0(I))$. Hence, the number of susceptible becomes a function of the number of infected.

\paragraph{SIR model.}

Under the assumption that the number of susceptible individuals depends on the number of infected through the initial viral load, we obtain a closed equation for the number of infected:
\begin{equation}\label{c1}
  \frac{dI}{dt} = \lambda I S(v_0(I)) - \nu I .
\end{equation}
As before, let us assume that the total population consists of two sub-populations, one of them with weak immune response and another one with strong immune response. If the initial viral load $v_0$ is less than some critical value $\hat v_0$, then the maximal viral concentration in the case of strong immune response remains less than the critical level $v_c$ which determines the appearance of the disease. This critical viral level can be determined from the equality $v_c = \hat v_0 \exp(W/s_2)$. Hence

$$ S(v_0) = \left\{
\begin{array}{cc}
S_1 , & v_0 < \hat v_0 \\
S_2 ,& v_0 \geq \hat v_0
\end{array} \right. \; , $$
where $S_1$ corresponds to the sub-population with weak immune response and $S_2$ to the total population.
Under this assumption, the explicit dependence $v_0(I)$ is not needed. We suppose that the viral load is an increasing function such that $v_0(I) < \hat v_0$ for $I<I_c$, and $v_0(I) \geq \hat v_0$ for $I \geq I_c$. Then instead of equation (\ref{c1}) we can write:
\begin{equation}\label{c2}
  \frac{dI}{dt} = \lambda I S_1 - \nu I \; , \;\; I < I_c \; ; \;\;\;
  \frac{dI}{dt} = \lambda I S_2 - \nu I \; , \;\; I \geq I_c \; .
\end{equation}
We can now find the solution:
$$ I(t) = I_0 e^{\alpha_1 t} \; , \;\; 0 < t < t_c= \frac{1}{\alpha_1} \ln\left(\frac{I_c}{I_0}\right) \; , \;\;\; I(t) = I_c e^{\alpha_2 (t-t_c)} \; , \;\; t_c < t  \; . $$
Here $\alpha_1=\lambda S_1-\nu$, $\alpha_2=\lambda S_2-\nu$. Hence, the number of infected individuals grows with exponent $\alpha_1$ till some critical time $t_c$, and with a larger exponent $\alpha_2$ for $t > t_c$.

\paragraph{Quarantine model.}

Another epidemiological model was introduced in \cite{VBP}, in order to take into account removal of infected individuals to quarantine when the incubation period is finished and they manifest disease symptoms:
\begin{equation}\label{c3}
  \frac{dI}{dt} = \lambda I(t) S(t) - \lambda I(t-\tau) S(t-\tau) .
\end{equation}
Here $I(t)$ is the density of latently infected individuals (during the incubation period), $\tau$ is the length of the incubation period. In this case we have three time intervals: $0<t<t_c$ where $t_c$ is such that $I(t_c)=I_c$; $t_c < t < t_c+\tau$, and $t>t_c+\tau$. We assume $S(t)=S_1$, $S(t-\tau)=S_1$ for $0<t<t_c$, and the solution of equation (\ref{c3}) is as follows:
$$ I(t) = I_0 e^{\mu_1 t} \; , \;\; 0 < t < t_c , $$
where $I_0$ is the initial number of latently infected individuals, and $\mu_1$ satisfies the equation $\mu_1 = kS_1 (1-e^{-\mu_1 \tau})$. This equation has a positive solution if the basic reproduction number $\mathcal{R}_1 = \lambda S_1 \tau$ is greater than $1$. Let us recall that $\mathcal{R}_1$ characterizes the number of newly infected individuals during the incubation period. The value $t_c$ can be found from the equality $I(t_c)=I_c$, $t_c = \frac{1}{\mu_1}\ln\left(\frac{I_c}{I_0}\right)$.

During the second time interval $t_c < t < t_c+\tau$, we have $S(t)=S_2$ and $S(t-\tau)=S_1$. The solution of equation (\ref{c3}) writes:
$$ I(t) = I_c e^{\mu_2 (t-t_c)} \; , \;\; t_c < t < t_c + \tau , $$
where $\mu_2$ satisfies the equality $\mu_2 = \lambda S_2 (1- \theta e^{-\mu_2 \tau})$, $\theta = S_1/S_2$. Since $\theta<1$, this equation has a positive solution and hence $\mu_2>0$.

If $t > t_c + \tau$, then $S(t)=S_2$, $S(t-\tau)=S_2$,
$$ I(t) = I_1 e^{\mu_3 (t-t_c-\tau)} \; , \;\;  t > t_c+\tau, $$
where $I_1=I_c e^{\mu_2 \tau}$, and $\mu_3$ satisfies the equation $\mu_3 = kS_2 (1-e^{-\mu_3 \tau})$. This equation has a positive solution if the basic reproduction number $\mathcal{R}_2 = \lambda S_2 \tau$ is larger than $1$.

Let us note that $\mu_1 < \mu_3 < \mu_2$. Therefore, the final growth rate is larger than the initial one but less than the intermediate.


\bigskip


\section{Discussion}

Numerous data on seasonal influenza epidemic shows that growth of the number of infected individuals can occur in two stages. Both of them are characterized by an exponential growth rate but the second exponent is larger than the first one. Growth accelerates when the number of infected reaches certain threshold. It should be noted that usually several virus strains are observed during influenza epidemic. All of them show similar growth pattern.

Such dynamics can not be captured by conventional epidemiological models like SIR (susceptible-infected-recovered). These models predict exponential growth at the beginning of epidemic when the number of susceptible $S$ can be considered as approximately constant or too large compared to initial number of infected. During the disease progression, $S$ decreases, as a consequence the growth rate of disease progression also decreases. The number of infected passes through a maximum and then decays to $0$.

The goal of this work is to develop a minimal mathematical model which can describe accelerated growth of disease spread. Our main hypothesis is that this acceleration can occur if we take into account the presence of different sub-populations, namely, with weak and strong immune response. In the beginning of epidemic, only (or mainly) people with weak immune response are concerned while at the second stage people with strong immune response also become susceptible. This transition takes place when the number of infected individuals becomes significantly large, and an uninfected individual has enhanced probability of interactions with infected individuals. This increases the viral load received during a short period of time and leads to the disease propagation in the sub-population with strong immune response.

In order to analyze how the initial viral load influences the disease initiation and progression, we consider an immuno-epidemiological model describing the concentration of viruses and immune cells in the organism. The strength of immune response in the model is determined by the rate of multiplication of immune cells. We show that for the same initial viral load, an individual with a weak immune system falls ill while an individual with a strong immune system does not develop disease symptoms. However, if the initial viral load increases, then the individuals with strong immune response also start developing the disease.

The combination of immunological and epidemiological  models allows the description of the interplay between the population level and individual level \cite{cai, gand, luk, pug, dorp}. We use here this approach to study a possible acceleration of epidemic growth rate.

We describe a generic situation applicable to different types of epidemics. During the influenza epidemic, the critical level of incidences when the growth accelerates is from 50 to 150 per week over 100 000 of the population (Figures \ref{flu}, \ref{UK}, \ref{Belgium}). Taking into account that the disease duration is about one week, then we get an estimate of the number of simultaneously infected individuals. The situation is different for the coronavirus infection because the disease duration is longer (about 4 weeks) and because the patients with manifested symptoms are put in quarantine. Therefore, we need to take into account here only the latently infected individuals during the incubation period. For a country with the total population 10 million and duration of the incubation period is one week, the critical level of epidemic is about 1400 new daily cases. This estimate does not take into account the restrictions on the population movement (it does not affect the influenza epidemic).

The immuno-epidemiological model developed in this work is not pathogen specific. We suppose that it can be also applied to coronavirus infection since the strength of adaptive immune response plays important role in the disease progression \cite{li}. It should be noted that in some cases excessive immune response can damage lung tissue and provoke important clinical consequences.

\paragraph{Limitations of the model.}

The main hypothesis of this work is that epidemic progression begins with a sub-population characterized by weak immune response and, at a later stage, it continues in the whole population. Though this assumption seems plausible, we do not have direct confirmation that this heterogeneity of the population plays a significant role during epidemic spread.

A particular choice of the epidemiological model is not essential here since all of them give exponential growth for the number of infected individuals. The exponential growth rate is proportional to the density of susceptible individuals. Once we suppose that effective density of susceptible can increase during epidemic because different sub-populations are involved one after another, we can obtain the increase of the growth rate.

The immunological model used in this study is very simplified. It does not take into account for the involvement of many different cell types, intracellular regulation, time delay in virus production and clonal expansion of immune cells, and some other relevant aspects. However, it captures the main features of the interaction of viral infection with the immune response. Its simplicity allows to couple it with the epidemiological model.

Let us note that system (\ref{b1}), (\ref{b2}) can also be explicitly solved if the functions $p(v)$ and $h(v)$ are linear (and the second assumption is satisfied).


\section*{Acknowledgements}

The last author acknowledges the IHES visiting program during which this work was done. The work was supported by the Ministry of Science and Education of Russian Federation, project number FSSF-2020-0018, and by the French-Russian program PRC2307.




\section{{\bf Appendix 1.} Influenza epidemic data}

\begin{figure}[ht!]
\centerline{\includegraphics[scale=0.2]{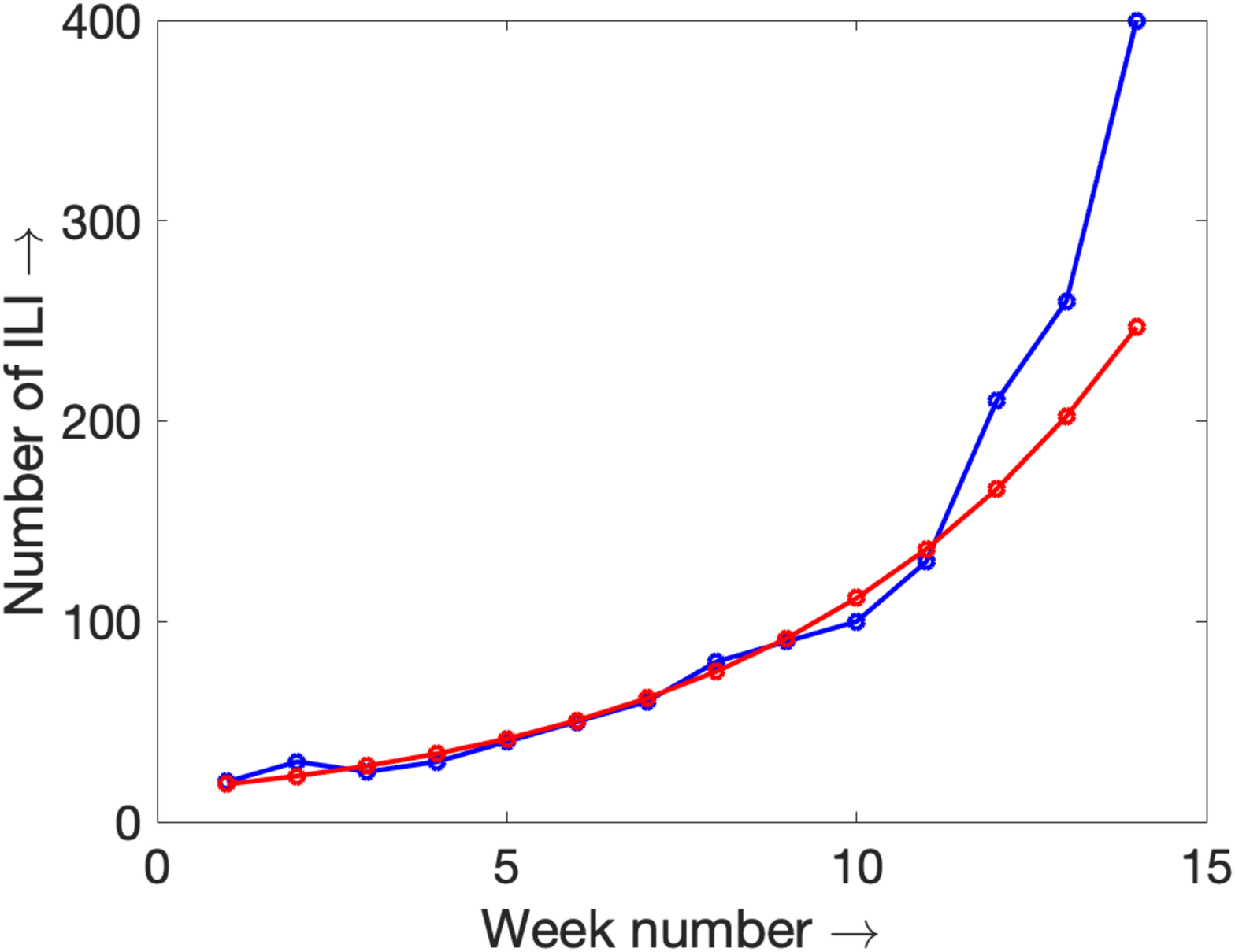}
\includegraphics[scale=0.2]{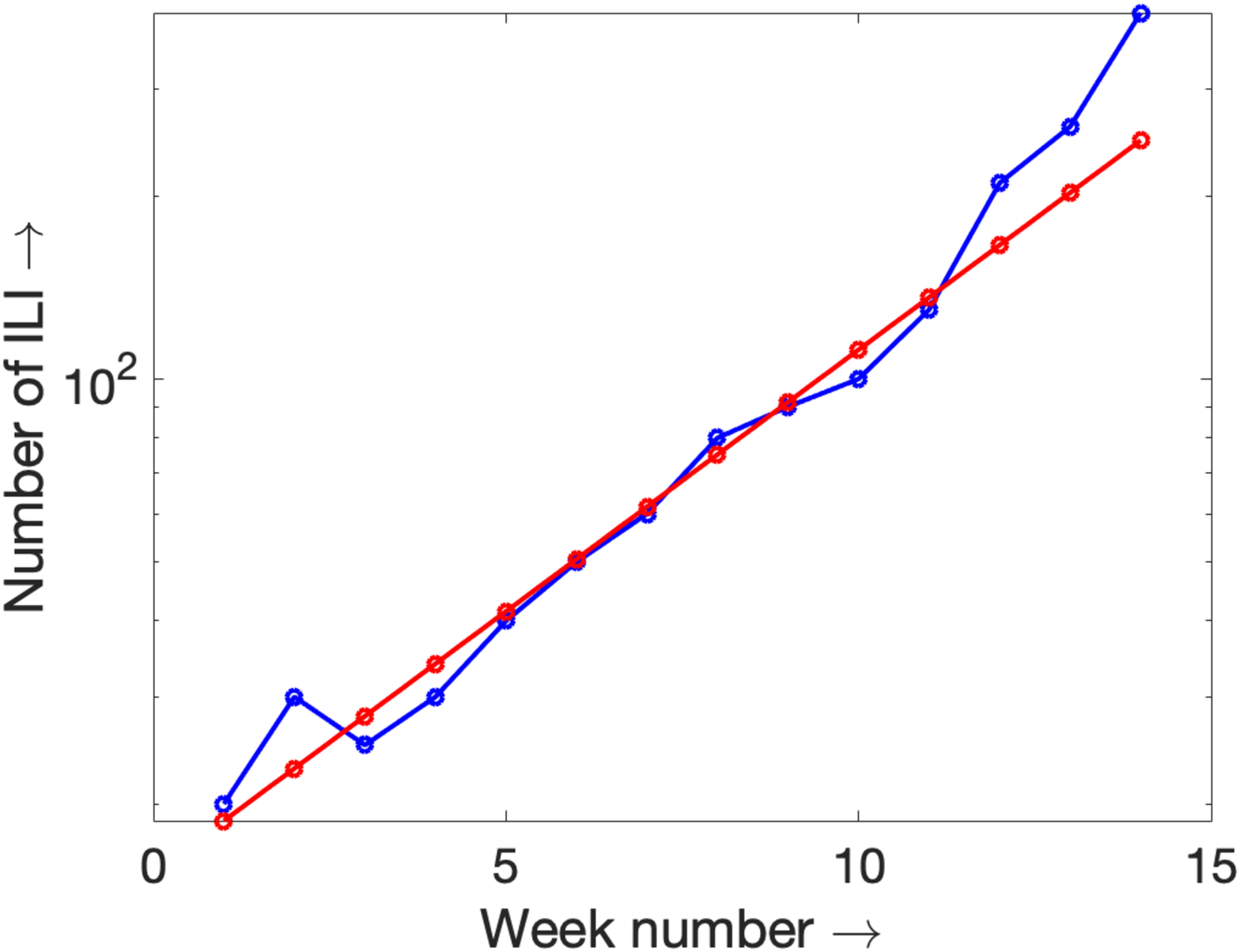}}
\caption{Data on seasonal influenza in UK \cite{uk}. The left graph shows the number of influenza-positive specimens from non-sentinel sources (blue line) and its approximation by an exponential (red line). The right graph shows the same curves in the logarithmic scale.}
\label{UK}
\end{figure}

\begin{figure}[ht!]
\centerline{\includegraphics[scale=0.2]{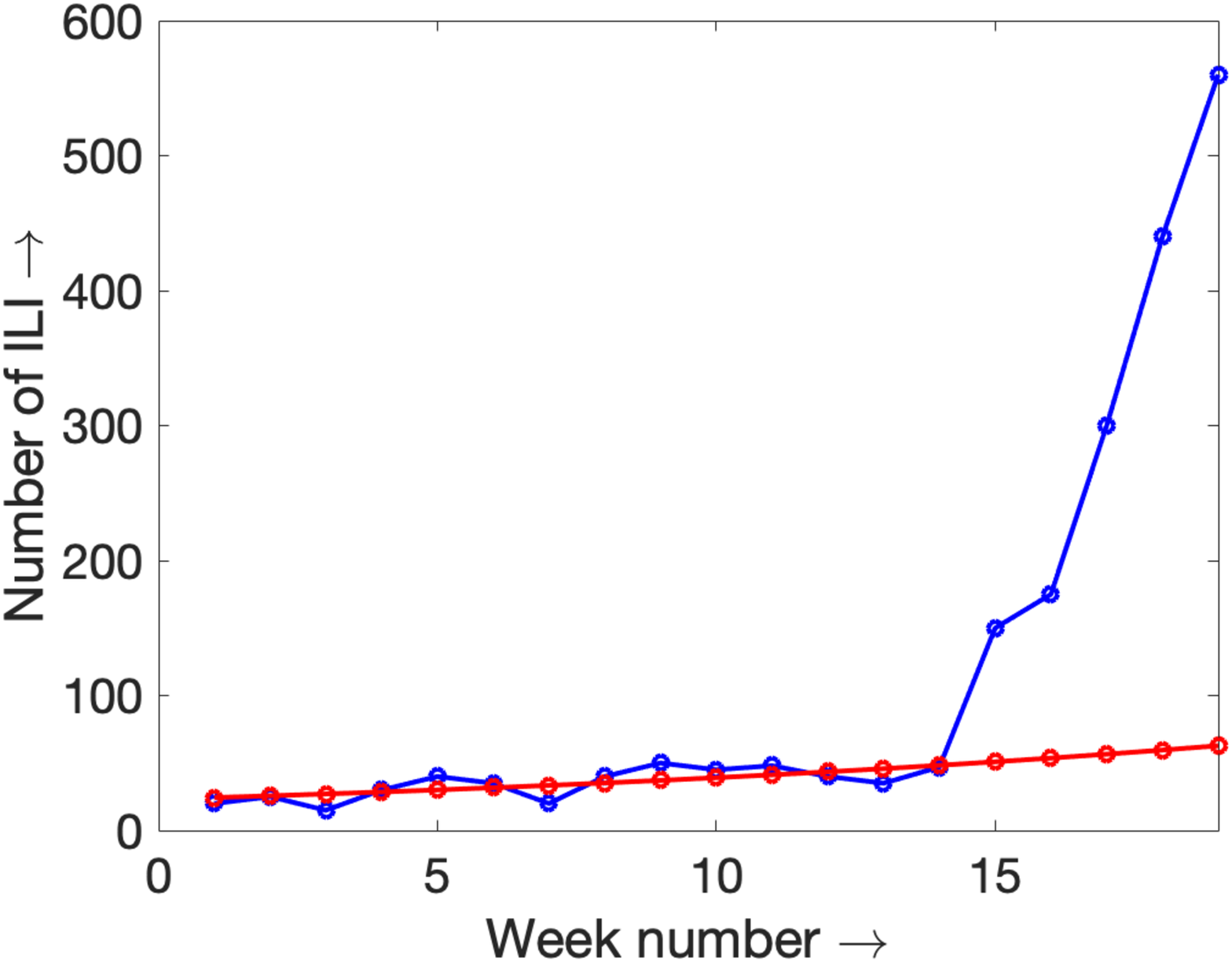}
\includegraphics[scale=0.2]{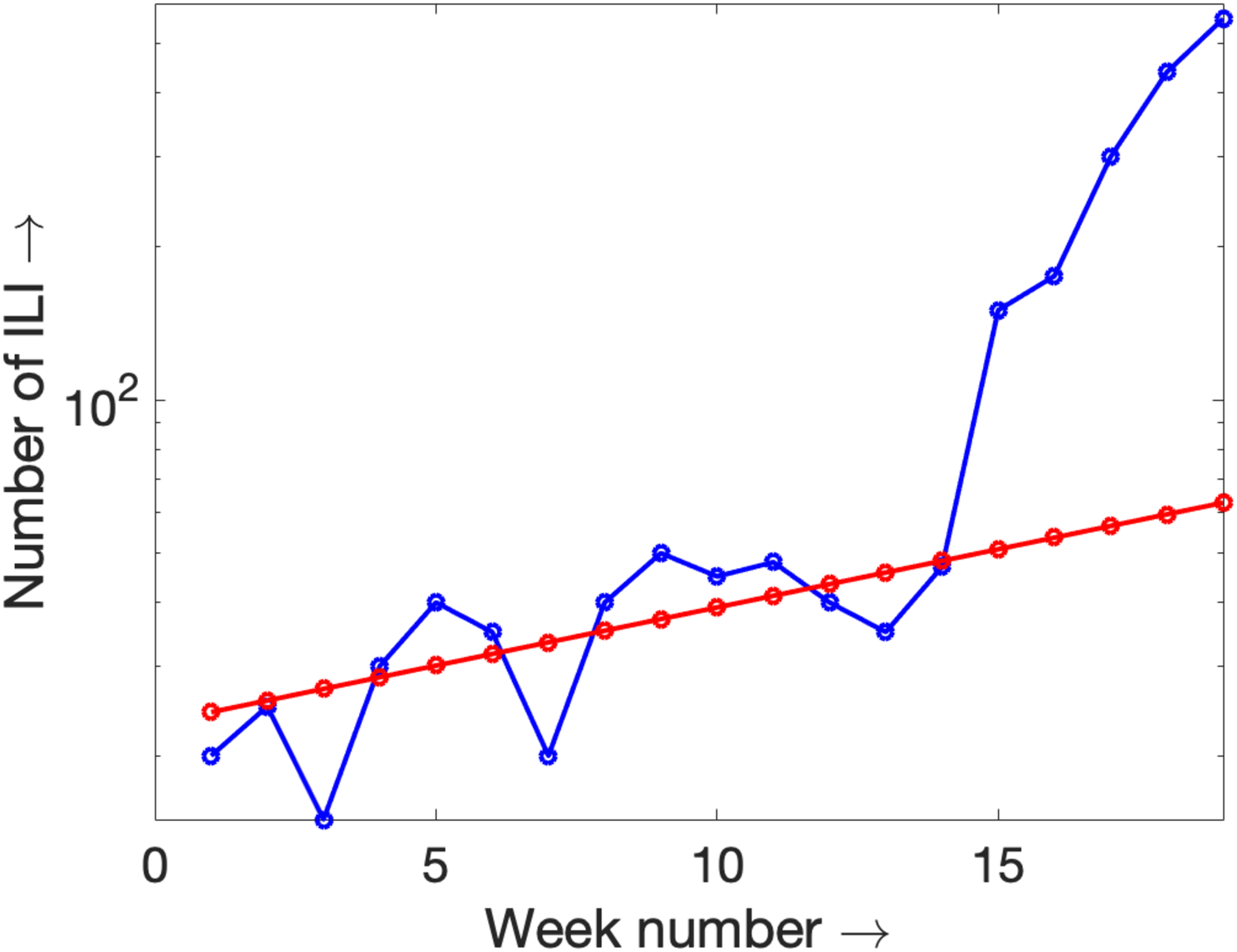}}
\caption{Data on seasonal influenza in Belgium \cite{belgium}. The left graph shows the number of influenza-positive specimens from non-sentinel sources (blue line) and its approximation by an exponential (red line). The right graph shows the same curves in the logarithmic scale.}
\label{Belgium}
\end{figure}

We present here some typical data on influenza epidemic (Figures \ref{UK}, \ref{Belgium}). The two stage growth can be observed.

\section{{\bf Appendix 2.} Solution of the immunological system}

\begin{figure}[ht!]
\centerline{\includegraphics[height=5cm,width=17cm]{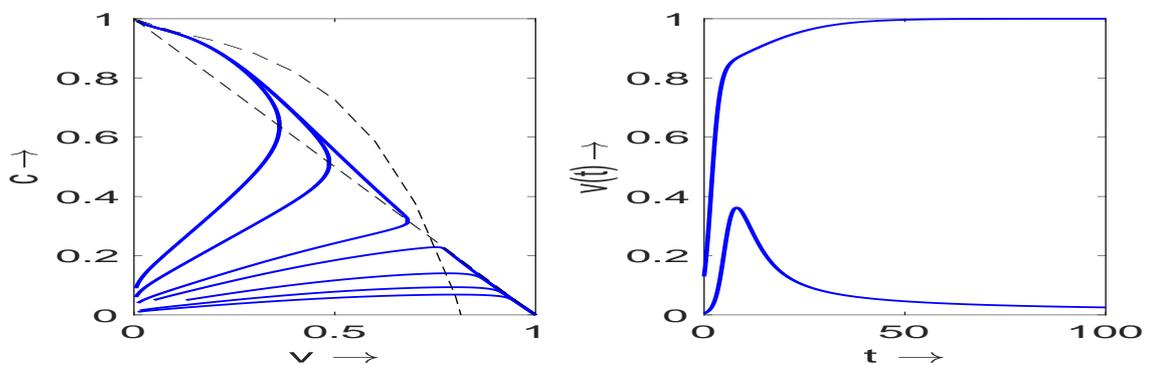}}
\caption{Trajectories on the phase plane of system (\ref{b1}), (\ref{b2}) in the case of weak immune response (left) and two sample plots for the growth of $v(t)$ with time (right).}
\label{traj1}
\end{figure}

\begin{figure}[ht!]
\centerline{\includegraphics[height=5cm,width=17cm]{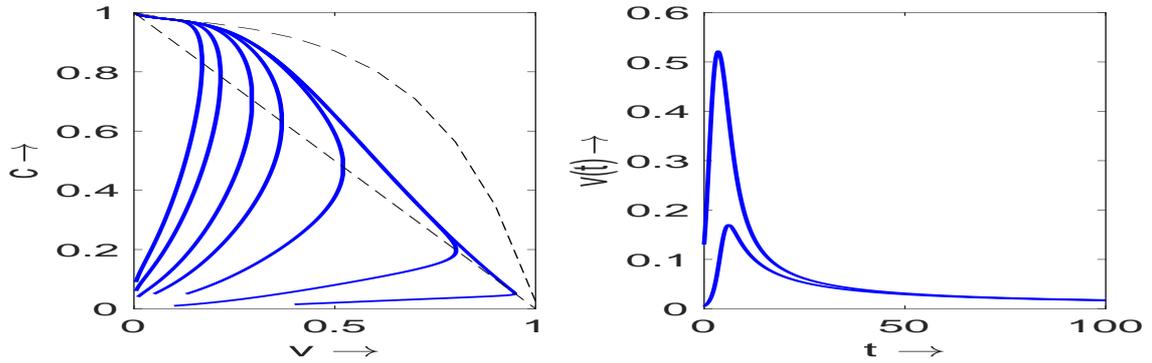}}
\caption{Trajectories on the phase plane of system (\ref{b1}), (\ref{b2}) in the case of strong immune response (left) and two sample plots for the growth of $v(t)$ with time (right).}
\label{traj2}
\end{figure}

We present here some results of numerical simulations of system (\ref{b1}), (\ref{b2}) with the functions $p(v) = (p_0 + p_1v)e^{-p_3v}$, $h(v) = h_1 v$, where $p_0=0.2$, $p_1=5$ (Figure \ref{traj1}) and $p_1=10$ (Figure \ref{traj2}), $p_3=4, h_1=0.2$. Other two parameter values are $k=1$ and $\sigma=1$.

The strength of immune response is characterized by the parameter $p_1$ which determines the rate of clonal expansion of immune cells in response to the antigen. A lesser value of $p_1$ corresponds to weaker immune response (Figure \ref{traj1}), a larger value of $p_1$ to stronger immune response (Figure \ref{traj2}).

There are 7 trajectories in each figure starting from the same (pairwise) initial condition for some small values $v=v_0$ (initial viral load) and $c=c_0$ (initial concentration of antigen specific cells). If the initial viral load is sufficiently small, then in both cases the trajectories converge to the virus-free stationary point corresponding to the curing. However, in the case of strong immune response the maximal and the total virus concentrations are less. Therefore, strong immune response can prevent the development of the disease while weak immune response may not be sufficient for this. If the initial viral load is sufficiently large, then weak immune response can lead to a chronic infection or to death while strong immune response leads to virus eradication.

\end{document}